# Observables in non-Hermitian systems: A methodological comparison

Karin Sim[1,*] Nicolò Defenu[1], Paolo Molignini[2], and R. Chitra[1]
[1]*Institute for Theoretical Physics, ETH Zürich, 8093 Zurich, Switzerland*
[2]*Department of Physics, Stockholm University, AlbaNova University Center, 106 91 Stockholm, Sweden*

Despite acute interest in the dynamics of non-Hermitian systems, there is a lack of consensus in the mathematical formulation of non-Hermitian quantum mechanics in the community. Different methodologies are used in the literature to study non-Hermitian dynamics. This ranges from consistent frameworks like biorthogonal quantum mechanics and metric approach characterized by modified inner products, to normalization by time-dependent norms inspired by open quantum systems. In this work, we systematically explore the similarities and differences among these various methods. Utilizing illustrative models with exact solutions, we demonstrate that these methods produce not only quantitatively different results but also distinct physical interpretations. For dissipative systems where non-Hermiticity arises as an approximation, we find that simply dividing by the norm in the $\mathcal{PT}$-broken regime closely aligns with the full master equation solutions. In contrast, for quantum systems where non-Hermiticity can be engineered exactly, incorporating metric dynamics is crucial for the probabilistic interpretation of quantum mechanics, necessitating the generalizations of similarity transformations and unitarity to non-Hermitian systems. This study lays the groundwork for further exploration of non-Hermitian Hamiltonians, potentially leveraging generalized transformations for novel physical phenomena.

## I. INTRODUCTION

Systems governed by non-Hermitian Hamiltonians [1–3] serve as an ideal playground to explore a wide range of unconventional behaviors. Examples include exceptional points [4], unconventional topology [5–7], and quantum phase transitions without gap closure [8]. Non-Hermiticity arises in many different contexts. For example, it provides an effective description of Lindbladian dynamics under certain conditions in dissipative systems [9] and continuously monitored systems [10,11]. Non-Hermitian Hamiltonians can also be directly engineered in coupled qubit systems via Naimark dilation [12–14]. Recently, numerous experimental implementations of non-Hermitian systems have emerged, proving that they are far from merely theoretical endeavors. Examples include optical [15] and acoustic [16] systems, ultracold atoms [17], and quantum circuits [12].

Non-Hermitian Hamiltonians which commute with the $\mathcal{PT}$ operator often possess a real spectrum in the regime where $\mathcal{PT}$ symmetry is not spontaneously broken [3,18,19]. $\mathcal{PT}$ symmetry [3,20] refers to the combined operation of parity and time reversal symmetries. This has prompted the interpretation of this special class of non-Hermitian systems as a natural extension to conventional quantum mechanics [21]. When $\mathcal{PT}$ symmetry is spontaneously broken, exceptional points (EPs) [22] arise in the spectrum, where the eigenvectors of the Hamiltonian coalesce and the spectrum becomes gapless. Beyond the EP, the spectrum becomes complex. Exceptional points (EPs) have garnered significant recent attention from both theoretical [2,4,23] and experimental [24,25] perspectives.

Dynamics of non-Hermitian systems is a new frontier of exploration which harbors a lot of potential to discover new physical paradigms. A plethora of novel phenomena, such as dynamically induced topological phases, exponentially enhanced quantum sensing [26] and the divergence of Lieb-Robinson bounds [8,27] have been reported recently. Despite its increasing popularity, there is a lack of consensus in the mathematical formulation of non-Hermitian quantum mechanics in the community. As non-Hermitian time evolution is intrinsically nonunitary, different methods have been adopted to obtain a normalized physical observable in non-Hermitian systems. Among them is biorthogonal quantum mechanics (BQM) [28], which is a comprehensive formulation of non-Hermitian quantum mechanics consistent with its probabilistic interpretation. BQM is widely adopted within the condensed matter community [6,29–31] and mainly applied to study systems featuring a time-independent Hamiltonian. The so-called metric framework [32–35] provides a generalization of BQM to the case with time-dependent Hamiltonians. In this framework, one considers a modified Hilbert space endowed by a time-dependent metric which appears as a weight factor in the inner product. The equation of motion of the metric incorporates the dynamics induced by both the Hamiltonian and its Hermitian conjugate. Both of these methods generate an exact time evolution of the non-Hermitian system, and are best suited when the non-Hermiticity of the system is realized in a fundamental manner instead of arising as an approximation. For example, this can be realized in NV centres via Naimark dilation [12].

---

*Contact author: simkarin@phys.ethz.ch

On the other hand, a lot of the available literature on non-Hermitian dynamics [27,36–40] skirts the important issue of nonconserved norms during time evolution and simply divides physical quantities by the time-dependent norm of the time-evolved state. This approach is ubiquitous in the open quantum systems community and is motivated by the quantum trajectory theory [41–45]. Quantum trajectories are used to find the time evolution of open quantum systems [46], for example when the direct solution of Lindblad master equation is computationally costly. In certain scenarios such as the detection of photons emitted from a cavity, the evolution of the system can be approximated by that with an effective non-Hermitian Hamiltonian interrupted by quantum jumps at random times [41]. By post-selecting quantum trajectories which did not undergo quantum jumps, a non-Hermitian evolution can be realized and the time-evolved state has to be explicitly normalized by hand at every time step [36].

The ostensibly different methodologies lead to very different physics. A prominent example is the quantum Brachistochrone problem [47,48]. In this problem, the division-by-norm method predicts arbitrarily fast quantum evolutions using $\mathcal{PT}$ symmetry [47]. However, Ref. [48] shows, using the metric formalism, that it is impossible to achieve faster unitary evolutions using non-Hermitian Hamiltonians than those given by Hermitian Hamiltonians. Another example is our recent finding [49] that the metric formalism predicted nonzero defect density due to $\mathcal{PT}$-broken modes when crossing exceptional points adiabatically. This violation of adiabaticity was deeply linked to the even parity of the observables with respect to momentum, an aspect absent in the division-by-norm methods.

These puzzling discrepancies motivate an in-depth comparative study of the properties of observables in non-Hermitian systems from a methodological perspective. In this work, we clarify the differences and similarities between the various methodologies used to study non-Hermitian dynamics and their physical impact. More importantly, we show that the consistent non-Hermitian quantum mechanical formulations, biorthogonal and metric, necessitate a generalization of the notions of similarity transformations and unitarity in non-Hermitian settings. Our work highlights the physical origin of the structures of the observables with respect to momentum, seen in Ref. [49], which arises naturally in the metric formalism.

The paper is structured as follows. We first introduce the mathematical background of the different methodologies in Sec. II, then proceed in Sec. III to show the equivalence of BQM, pseudo-Hermiticity, and the metric framework when the Hamiltonian is time-independent and has a real spectrum. In Sec. IV, we study the dynamics of open quantum systems, and compare the results of the metric framework, no-jump approximation, and the full master equation. In Sec. V, we generalize the notion of unitarity in the context of biorthogonal framework and elaborate on its physical consequence through similarity transformations of the Hamiltonian which maps between different momentum modes.

## II. METHODOLOGIES

Here, we briefly describe the different approaches that are used in the study of non-Hermitian systems. We first present the consistent formulations of non-Hermitian quantum mechanics: biorthogonal and metric frameworks, followed by the approaches used in the case of dissipative quantum systems.

### A. Biorthogonal quantum mechanics (BQM)

We start with a brief summary of BQM [28] for a general static non-Hermitian Hamiltonian $H$ with a nondegenerate spectrum. We assume a finite-dimensional Hilbert space throughout the paper. The infinite-dimensional case has been discussed in Refs. [28,50]. The left and right eigenstates of $H$ are defined as

$$H|n\rangle_R = E_n|n\rangle_R,$$
$$H^\dagger|n\rangle_L = \overline{E}_n|n\rangle_L, \quad (1)$$

where the sets $\{|n\rangle_R\}$ and $\{|n\rangle_L\}$ individually form a complete basis and are linearly independent, though not necessarily orthogonal [28]. The left and right eigenstates satisfy

$$_L\langle m|n\rangle_R = \delta_{mn}. \quad (2)$$

The identity operator can now be expressed in terms of both left and right eigenstates as

$$\sum_m |m\rangle_{RL}\langle m| = \mathbb{1}. \quad (3)$$

Given a general state $|\psi\rangle$, we can decompose it in the right eigenbasis as

$$|\psi\rangle = \sum_n c_n |n\rangle_R, \quad (4)$$

where $c_n = {}_L\langle n|\psi\rangle$. For the non-Hermitian case, we introduce the *associated state*

$$|\tilde{\psi}\rangle := \sum_n c_n |n\rangle_L, \quad (5)$$

as expectation values of observables in the biorthogonal framework are defined as $\langle \hat{F} \rangle = \langle \tilde{\psi}|\hat{F}|\psi\rangle$. In particular, setting $\hat{F} = \mathbb{1}$, we see $\langle \tilde{\psi}|\psi\rangle = \sum_n |c_n|^2 = 1$.

The notion of Hermiticity changes in the context of BQM. This can be seen by considering the expansion of an observable $\hat{F}$ in the biorthogonal basis

$$\hat{F} = \sum_{mn} f_{mn} |m\rangle_{RL}\langle n|, \quad (6)$$

from which it can be seen that the realness of the expectation value $\langle \tilde{\psi}|\hat{F}|\psi\rangle = \langle \psi|\hat{F}^\dagger|\tilde{\psi}\rangle$ is only guaranteed if $f_{mn} = f^*_{nm}$; this defines *biorthogonally Hermitian* operators [28].

### B. Metric framework

An alternative to the BQM formalism is the metric approach to non-Hermitian quantum mechanics [33,34,51,52]. Here, we consider a modified Hilbert space endowed with a positive-definite "metric" operator $\rho(t)$, which appears as a weight factor in the inner product as $\langle \cdot, \cdot \rangle_{\rho(t)} := \langle \cdot |\rho(t)|\cdot\rangle$.

The time evolution of $\rho(t)$ follows from the requirement that the norm of the time-evolved state is conserved. Its equation of motion is given by

$$i\dot{\rho}(t) = H^\dagger(t)\rho(t) - \rho(t)H(t), \qquad (7)$$

where the overdot denotes time derivative and $H(t)$ is a non-Hermitian Hamiltonian. Here we generalize to include the case of a time-dependent Hamiltonian, where this formalism is also valid. Additionally, dynamics is an integral part of this framework, distinguishing it from the earlier discussion of BQM which is more naturally applied to a stationary Hamiltonian. Using the solution to Eq. (7) and the square-root decomposition $\rho(t) = \eta^\dagger(t)\eta(t)$, we can map the system to an effective Hermitian Hamiltonian

$$h(t) = \eta(t)H(t)\eta^{-1}(t) + i\dot{\eta}(t)\eta^{-1}(t). \qquad (8)$$

We assume $\eta(t)$ to be Hermitian, since all non-Hermitian square roots are related to the Hermitian root by similarity transformations; see, e.g., Ref. [49] for a discussion on this. The Hamiltonian $h(t)$ acts in a different Hilbert space $\mathscr{H}$ where the non-Hermiticity is encoded in the dynamics of $\eta(t)$.

Time evolution in the Hilbert spaces $\mathscr{H}_{\rho(t)}$ and $\mathscr{H}$ is generated by the respective Hamiltonians, $H(t)$ and $h(t)$, via time-dependent Schrödinger equation

$$\begin{aligned} i\frac{d}{dt}|\psi(t)\rangle &= H(t)|\psi(t)\rangle, \\ i\frac{d}{dt}|\Psi(t)\rangle &= h(t)|\Psi(t)\rangle, \end{aligned} \qquad (9)$$

where the states are related by $|\Psi(t)\rangle = \eta(t)|\psi(t)\rangle$. The construction of $\rho(t)$ guarantees that $h(t)$ is Hermitian, such that the unitarity of the time evolution is restored. In particular, Eq. (7) guarantees that $\langle\psi(t)|\rho(t)|\psi(t)\rangle = \langle\psi(0)|\rho(0)|\psi(0)\rangle$ at all times $t$. A natural choice of initial condition is then $\rho(0) = \mathbb{1}$ [49], which ensures that $\langle\psi(t)|\rho(t)|\psi(t)\rangle = \langle\Psi(t)|\Psi(t)\rangle = 1$ given a normalized initial state $|\psi(0)\rangle = |\Psi(0)\rangle$.

The expectation value of an operator $\hat{o}: \mathscr{H} \to \mathscr{H}$ is given by

$$\begin{aligned} \langle\hat{O}(t)\rangle_{\text{metric}} &= \langle\Psi(t)|\hat{o}|\Psi(t)\rangle \\ &= \langle\psi(t)|\rho(t)\hat{O}(t)|\psi(t)\rangle, \end{aligned} \qquad (10)$$

where $\hat{O}(t): \mathscr{H}_{\rho(t)} \to \mathscr{H}_{\rho(t)}$ is defined as $\hat{O}(t) = \eta^{-1}(t)\hat{o}\eta(t)$.

### C. Open quantum systems

As dissipation is a common source of non-Hermiticity in quantum systems[41–43,46], we provide a brief recap of the theory underlying such systems. Within the Born-Markov approximations, the dynamics of an open quantum system is well-approximated by the Born-Markov master equation [53], which can be written in the Lindblad form as follows:

$$\begin{aligned} \dot{\rho}_{\text{me}} = &-i[H, \rho_{\text{me}}] + \sum_j \Gamma_j \rho_{\text{me}} \Gamma_j^\dagger \\ &- \sum_j \frac{1}{2}\{\rho_{\text{me}}, \Gamma_j^\dagger \Gamma_j\}. \end{aligned} \qquad (11)$$

The master equation describes the time evolution of the system density operator $\rho_{\text{me}}$, which represents a probabilistic mixture of pure states. In Eq. (11), $H$ is the system Hamiltonian and each $\Gamma_j$ is a jump or Lindblad operator, which encodes the interaction between the system and the environment. Here, the Hamiltonian $H$ can be time-dependent, but we assume the Lindblad operators to be time-independent. The expectation value of an observable $\hat{O}$ in such a system is defined as $\langle\hat{O}\rangle_{\text{ME}} = \text{Tr}(\rho_{\text{me}}\hat{O})$. For simplicity, we take the case with only one Lindblad operator, i.e., $\Gamma_j = \Gamma$ for $j = 1$ and 0 otherwise. The result is readily generalizable to the case with multiple Lindblad operators.

The direct solution of Eq. (11) is often computationally costly, and the quantum trajectory approach [42] is used as an alternative in such a case. The method entails computing the stochastic evolution of the individual pure states in the ensemble, which correspond to the states conditioned on the measurement outcomes. Numerically, this involves the evolution of a pure state with an effective non-Hermitian Hamiltonian $H_{\text{eff}} = H - \frac{i}{2}\Gamma^\dagger\Gamma$, interrupted by finite evolutions induced by one of the measurement operators at random times. These are known as quantum jumps [42]. The ensemble average of the projectors onto the pure states then give rise to the density operator $\rho_{\text{me}}$, and the expectation values of observables can also be calculated accordingly. The evolution of the individual pure state is discontinuous in time, unlike the continuous evolution of the density matrix which represents the ensemble average.

The so-called no-jump limit [54,55] corresponds to the evolution of a dissipative system in the absence of quantum jumps. For example, by post-selecting quantum trajectories that did not undergo a jump, we can describe the evolution of the system purely in terms of $H_{\text{eff}}$. This is equivalent to rewriting Eq. (11) in the form

$$\dot{\rho}_{\text{me}} = -i(H_{\text{eff}}\rho_{\text{me}} - \rho_{\text{me}}H_{\text{eff}}^\dagger) + \Gamma\rho_{\text{me}}\Gamma^\dagger, \qquad (12)$$

and neglecting the term $\Gamma\rho_{\text{me}}\Gamma^\dagger$, which is known as the recycling or quantum jump term. Curiously, the approximate master equation in the no-jump limit, $\dot{\rho}_{\text{nj}} = -i(H_{\text{eff}}\rho_{\text{nj}} - \rho_{\text{nj}}H_{\text{eff}}^\dagger)$ is identical to the metric equation of motion Eq. (7) with the substitution $H = H_{\text{eff}}^\dagger$. Note however, that the initial conditions are different. For example, a natural choice of initial condition for the metric is $\rho(0) = \mathbb{1}$, as is discussed in Sec. III and Ref. [49]. On the other hand, the initial state $\rho_{\text{nj}}(0)$ is usually taken to be a density matrix. For example, the density matrix of a qubit always takes the form $\rho_{\text{nj}}(0) = \frac{1}{2}(\mathbb{1} + \mathbf{r}\cdot\boldsymbol{\sigma}), |\mathbf{r}| \leqslant 1$. Note that the time-evolved $\rho_{\text{nj}}(t)$ does not describe a mixed state and is not trace-preserving [54,55]. In the no-jump limit, the expectation value of an observable is defined as $\langle\hat{O}\rangle_{\text{nj}} = \text{Tr}(\rho_{\text{nj}}\hat{O})/\text{Tr}(\rho_{\text{nj}})$. If the initial state is a pure state, the solution to the no-jump master equation is given by $\rho_{\text{nj}}(t) = |\psi(t)\rangle\langle\psi(t)|$, where $i|\dot{\psi}(t)\rangle = H_{\text{eff}}|\psi(t)\rangle$. Consequently [27,36–40],

$$\langle\hat{O}(t)\rangle_{\text{nj}} = \frac{\langle\psi(t)|\hat{O}|\psi(t)\rangle}{\langle\psi(t)|\psi(t)\rangle}, \qquad (13)$$

which is the same as tracking the Schrödinger evolution of the pure state $|\psi(t)\rangle$ and normalizing quantities by the time-dependent norm, as was done in many works [27,36–40]. To summarize, we have presented the different methodologies for handling non-Hermitian Hamiltonians: the fully self-consistent formulation of quantum mechanics encoded by both BQM and the metric, and the division-by-norm method used in the case of dissipative systems.

## III. METRIC AS A GENERALIZATION OF BQM

In this section, we show explicitly that the metric framework is a generalization of the consistent non-Hermitian quantum mechanics encapsulated by BQM, introduced in Sec. II A. We first summarize the findings of Refs. [28,56], which show the equivalence between BQM and a particular similarity transformation known as pseudo-Hermiticity [56–58]. We then illustrate how the time-dependent metric extends the notion of pseudo-Hermiticity to dynamical cases [33]. This motivates the metric framework as a natural generalization of BQM to the dynamical context.

For clarity, we consider a general time-independent Hamiltonian $H$ with a real, nondegenerate spectrum and an appropriately normalized biorthogonal basis fulfilling Eqs. (2) and (3). For a Hamiltonian with degenerate spectrum, the left and right eigenstates may not form complete sets, in which case the similarity transformation given by Eq. (15) does not exist [28].

It has been shown that every Hamiltonian with a real spectrum is pseudo-Hermitian [57]. The definition of pseudo-Hermiticity is the existence of the similarity transformation

$$H^\dagger = SHS^{-1}. \quad (14)$$

In the biorthogonal basis, we can write $S$ and $S^{-1}$ as [28,56]

$$S = \sum_n |n\rangle_{LL}\langle n|, \quad S^{-1} = \sum_n |n\rangle_{RR}\langle n|, \quad (15)$$

such that $SS^{-1} = \mathbb{1}$ by Eqs. (2) and (3). We have $S|n\rangle_R = |n\rangle_L$ and $S^{-1}|n\rangle_L = |n\rangle_R$, and so $S|\psi\rangle = |\tilde\psi\rangle$ according to Eqs. (4) and (5). Furthermore, for any operator $F$, we have the equivalence [28]

$$\langle\psi|SF|\psi\rangle = \langle\tilde\psi|F|\psi\rangle. \quad (16)$$

This shows the equivalence of observables obtained using BQM and by considering the pseudo-Hermiticity transformation (15). We also note that $\langle\psi|S|\psi\rangle = \langle\tilde\psi|\psi\rangle = 1$.

We now introduce the Hermitian square root of $S$, which we denote $\sqrt{S}$, such that $S = \sqrt{S}^2$. As was discussed in Sec. II B and Ref. [49], we assume this to be Hermitian. Defining the state $|\Psi\rangle = \sqrt{S}|\psi\rangle$, we see that

$$\langle\Psi|F|\Psi\rangle = \langle\psi|S\tilde F|\psi\rangle = \langle\tilde\psi|\tilde F|\psi\rangle, \quad (17)$$

where $\tilde F = \sqrt{S}^{-1}F\sqrt{S}$.

Equation (17) closely resembles the equivalence of the Hilbert spaces $\mathscr{H}$ and $\mathscr{H}_{\rho(t)}$ in the metric framework, in the particular case where the metric is time-independent. This is clear when we identify $\sqrt{S} \to \eta$, $\tilde F \to H$ and $F \to h$ in Eqs. (7) and (8). In fact, the matrix $S$, which satisfies Eq. (14), is a time-independent solution to the metric equation of motion Eq. (7) with the initial condition $\rho(0) = S$, such that $\dot\rho = 0$. A caveat is that not all non-Hermitian Hamiltonians can be mapped to its Hermitian conjugate by a similarity transformation. The existence of such a mapping requires the eigenvalues to be real and nondegenerate [51], so a time-independent solution $S$ to Eq. (7) does not generically exist. However, even when a stationary solution is not permissible, a dynamical solution $\rho(t)$ can be found. It is then evident that the notions in the metric framework are generalizations of the concepts discussed above, namely BQM and pseudo-Hermiticity, to time-dependent cases.

## IV. DYNAMICS OF NON-HERMITIAN SYSTEMS

With a view to understanding the physical content of the different formalisms discussed earlier, we provide a comprehensive comparison of the dynamics of observables stemming from the metric, the no-jump limit in a dissipative system and the full master equation. We do so through illustrative models of non-Hermiticity engendered by dissipation in two-level systems. We consider two representative cases (i) $[\Gamma^\dagger\Gamma, H] = 0$ and (ii) $[\Gamma^\dagger\Gamma, H] \neq 0$, where $H$ is the system Hamiltonian and $\Gamma$ is the Lindblad operator present in Eq. (11). Note that $H$ is Hermitian. We uncover behaviors that are sensitive to the underlying models and the initial conditions, but are starkly different in general.

### A. $\Gamma^\dagger\Gamma$ commutes with system Hamiltonian

We consider a dissipative system with system Hamiltonian $H = \omega\sigma_z$ and jump operator $\Gamma = 2\sqrt{\gamma}\sigma_- = \sqrt{\gamma}(\sigma_x - i\sigma_y)$, where $\omega, \gamma > 0$. For example, this corresponds physically to the decay process of a two-level system from the excited to the ground state. The effective non-Hermitian Hamiltonian, which appears in the no-jump master equation, is given by $H_{\text{eff}} = (\omega - i\gamma)\sigma_z$ up to a constant $i\gamma\mathbb{1}$ term. This term contributes to some factors in the time-evolved state and the metric. However, no matter which method we use (metric or division by norm), the factor cancels out when computing physical observables and therefore can be neglected. We note that the energy eigenvalues are always complex in this model.

The metric dynamics can be obtained analytically by solving Eq. (7) with $H = H_{\text{eff}}$. The solution is given by $\rho(t) = \cosh(2\gamma t)\mathbb{1} + \sinh(2\gamma t)\sigma_z$. The Hermitian mapping defined by Eq. (8), which acts in the Hilbert space $\mathscr{H}$, is given by $h = \omega\sigma_z$. This is a special case in which the Hermitian Hamiltonian $h$, the system Hamiltonian and the effective non-Hermitian Hamiltonian share an eigenbasis, which is the $\sigma_z$ basis denoted by $\{|0\rangle, |1\rangle\}$. For this model, the unique steady-state solution of the full master equation is given by $\lim_{t\to\infty}\rho_{\text{me}} = \frac{1}{2}(\mathbb{1} - \sigma_z) = |0\rangle\langle 0|$.

When the system is initially in the ground state of the system Hamiltonian, the $|0\rangle$ state, $\langle\hat O\rangle_{\text{metric}} = \langle\hat O\rangle_{\text{me}} = \langle\hat O\rangle_{\text{nj}} = \langle 0|\hat O|0\rangle$ is trivially the steady-state solution of the full master equation for a general observable $\hat O$. If we instead have the initial state $|\psi(0)\rangle = |1\rangle$, we have $\langle\hat O\rangle_{\text{metric}} = \langle\hat O\rangle_{\text{nj}} = \langle 1|\hat O|1\rangle$ which stays in the excited state. On the other hand, the solution of the full master equation

$$\langle\hat O\rangle_{\text{me}} = \langle 0|\hat O|0\rangle(1 - e^{-4\gamma t}) + e^{-4\gamma t}\langle\hat O\rangle_{\text{metric}}$$
$$= \langle 0|\hat O|0\rangle(1 - e^{-4\gamma t}) + e^{-4\gamma t}\langle\hat O\rangle_{\text{nj}} \quad (18)$$

approaches the steady-state solution in the long-time limit as expected.

However, for initial conditions that are superpositions of both the ground and excited states, we observe different behaviors. With the initial condition $|\psi(0)\rangle = \frac{1}{\sqrt{2}}(|0\rangle - |1\rangle)$, the metric observable oscillates indefinitely:

$$\langle \hat{O} \rangle_{\text{metric}} = \frac{1}{2}\text{Tr}(\hat{O}) - \cos(2\omega t)\text{Re}(\langle 1|\hat{O}|0\rangle) \\ + \sin(2\omega t)\text{Im}(\langle 1|\hat{O}|0\rangle). \quad (19)$$

On the other hand, the solution of the full master equation is given by

$$\langle \hat{O} \rangle_{\text{me}} = \frac{1}{2}\text{Tr}(\hat{O}) - \cos(2\omega t)e^{-2\gamma t}\text{Re}(\langle 1|\hat{O}|0\rangle) \\ + \sin(2\omega t)e^{-2\gamma t}\text{Im}(\langle 1|\hat{O}|0\rangle) \\ + \frac{e^{-4\gamma t} - 1}{2}(\langle 1|\hat{O}|1\rangle - \langle 0|\hat{O}|0\rangle), \quad (20)$$

which can be rewritten as

$$\langle \hat{O} \rangle_{\text{me}} - \frac{1}{2}\text{Tr}(\hat{O}) = \left(\langle \hat{O} \rangle_{\text{metric}} - \frac{1}{2}\text{Tr}(\hat{O})\right)e^{-2\gamma t} \\ + \frac{e^{-4\gamma t} - 1}{2}(\langle 1|\hat{O}|1\rangle - \langle 0|\hat{O}|0\rangle). \quad (21)$$

Comparing Eqs. (18) and (21), it is evident that while the master equation solution approaches the steady-state solution, the metric formalism predicts oscillatory behavior.

The solution to the no-jump master equation, or equivalently the norm approach, yields

$$\langle \hat{O} \rangle_{\text{nj}} = \frac{1}{2}\text{Tr}(\hat{O}) - \frac{2\cos(2\omega t)e^{-2\gamma t}}{1 + e^{-4\gamma t}}\text{Re}(\langle 1|\hat{O}|0\rangle) \\ + \frac{2\sin(2\omega t)e^{-2\gamma t}}{1 + e^{-4\gamma t}}\text{Im}(\langle 1|\hat{O}|0\rangle) \\ + \frac{e^{-4\gamma t} - 1}{2(e^{-4\gamma t} + 1)}(\langle 1|\hat{O}|1\rangle - \langle 0|\hat{O}|0\rangle). \quad (22)$$

This is comparable to the full master equation solution, up to some factors proportional to $\frac{1}{1+e^{-4\gamma t}}$ for each term except for the $\frac{1}{2}\text{Tr}(\hat{O})$ term. The factors arise due to renormalization by the trace of the approximate density matrix $\rho_{\text{nj}}$. In contrast to the metric solution, it tends to an asymptotic value that is equal to the steady-state solution in the long-time limit: $\lim_{t \to \infty}\langle \hat{O} \rangle_{\text{nj}} = \langle 0|\hat{O}|0\rangle$.

To summarize, the no-jump expectation value $\langle \hat{O} \rangle_{\text{nj}}$ tends to a constant value in asymptotically long time for general initial conditions away from the eigenbasis of $H_{\text{eff}}$. Even though $\langle \hat{O} \rangle_{\text{nj}}$ approaches the same value as the steady-state solution given by $\langle \hat{O} \rangle_{\text{me}}$, the short-time behaviors of the two expectation values only agree to zeroth order in time in general. On the other hand, the metric expectation value $\langle \hat{O} \rangle_{\text{metric}}$ is governed by the Hermitian mapping $h = \omega \sigma_z$ given by Eq. (8). Note that $h$ is independent of the non-Hermitian coefficient $\gamma$. As a result, $\langle \hat{O} \rangle_{\text{metric}}$ does not tend to an asymptotic value in the long-time limit for this model, independent of the initial condition.

### B. $\Gamma^\dagger \Gamma$ does not commute with system Hamiltonian

We now turn to a richer model in which the term $\Gamma^\dagger \Gamma$ does not commute with the system Hamiltonian. We set the system Hamiltonian to be $H = \omega \sigma_x$ and take the same jump operator as in the previous subsection, $\Gamma = 2\sqrt{\gamma}\sigma_- = \sqrt{\gamma}(\sigma_x - i\sigma_y)$, where $\omega, \gamma > 0$. For this model, the unique steady-state solution of the master equation is given by $\lim_{t\to\infty} \rho_{\text{me}} = \frac{1}{2}(\mathbb{1} + \mathbf{r}_{ss} \cdot \boldsymbol{\sigma})$, where

$$r_{x,ss} = 0, \quad r_{y,ss} = \frac{2\gamma\omega}{2\gamma^2 + \omega^2}, \quad r_{z,ss} = -\frac{2\gamma^2}{2\gamma^2 + \omega^2}. \quad (23)$$

The effective non-Hermitian Hamiltonian is $H_{\text{eff}} = \omega\sigma_x - i\gamma\sigma_z$ up to a term proportional to the identity operator. The exact solutions of Eqs. (8) and (9) are given in Appendix A. The energy eigenvalues are $\pm\sqrt{\omega^2 - \gamma^2}$. There are two parameter regimes: $\mathcal{PT}$-symmetric regime where $\gamma^2 \leqslant \omega^2$ and a regime where the $\mathcal{PT}$ symmetry is spontaneously broken, $\gamma^2 > \omega^2$. In the regime $\gamma^2 > \omega^2$, the energy eigenvalues are purely imaginary and appear as complex conjugate pairs. The significance of the different parameter regimes stems from the fact that the effective Hamiltonian $H_{\text{eff}}$ now features an exceptional point at $\omega = \gamma$. The solutions obtained by the no-jump approximation and the metric framework display parameter-dependent behaviors: in the $\mathcal{PT}$-broken regime $\omega^2 < \gamma^2$ shown by Figs. 1(b) and 1(d), the solutions tend to a constant value in the limit $t \to \infty$, whereas in the $\mathcal{PT}$-symmetric regime $\omega^2 > \gamma^2$ shown by Figs. 1(a) and 1(c), both methods yield oscillatory solutions. The only qualitative difference is that the metric expectation value displays periodic oscillations, while the oscillations of no-jump expectation value are aperiodic.

This parameter-dependent behavior is observed for various initial states as shown in Fig. 1. For the no-jump approximation, the asymptotic values valid for $\omega^2 < \gamma^2$ are given by

$$\lim_{t \to \infty}\langle \sigma_x \rangle_{\text{nj}} = 0, \quad \lim_{t \to \infty}\langle \sigma_y \rangle_{\text{nj}} = \frac{\omega}{\gamma},$$
$$\lim_{t \to \infty}\langle \sigma_z \rangle_{\text{nj}} = -\frac{\sqrt{\gamma^2 - \omega^2}}{\gamma}. \quad (24)$$

They are independent of initial conditions, since division by the time-dependent norm cancels out the contribution of the initial condition. An exception is when the initial state is an eigenbasis of $H_{\text{eff}}$, where the expectation values do not evolve in time. We note that the qualitative behaviors of $\langle \sigma_z \rangle_{\text{nj}}$ and $\langle \sigma_z \rangle_{\text{me}}$ are similar when $\omega^2 < \gamma^2$, though the asymptotic values are different. This can be seen from Figs. 1(b) and 1(d).

For the metric solutions, the asymptotic values obtained for the case $\omega^2 < \gamma^2$ are given by

$$\lim_{t \to \infty}\langle \sigma_x \rangle_{\text{metric}} = x_0,$$
$$\lim_{t \to \infty}\langle \sigma_y \rangle_{\text{metric}} = \frac{-2\omega\sqrt{\gamma^2 - \omega^2}z_0 + (\gamma^2 - 2\omega^2)y_0}{\gamma^2}, \quad (25)$$
$$\lim_{t \to \infty}\langle \sigma_z \rangle_{\text{metric}} = \frac{2\omega\sqrt{\gamma^2 - \omega^2}y_0 + (\gamma^2 - 2\omega^2)z_0}{\gamma^2}.$$

Unlike in the no-jump case, they depend on the initial conditions: $x_0$, $y_0$, and $z_0$ are the initial values of the spin expectation in each direction. The sensitivity to initial condition is typical of Schrödinger evolution of states, since the information of the initial state is encoded in the time-evolved state and thus in the expectation value. This emphasizes the fact that the metric framework tracks the Schrödinger dynamics of an intrinsically non-Hermitian system. The steady-state solution of the Markovian master equation is independent of the initial condition in general, since the presence of stochastic quantum jumps wipes out the information of the initial state in the long-time limit. Curiously, the no-jump approximation erases information about the initial state, though it describes the Hamiltonian dynamics of a pure state.

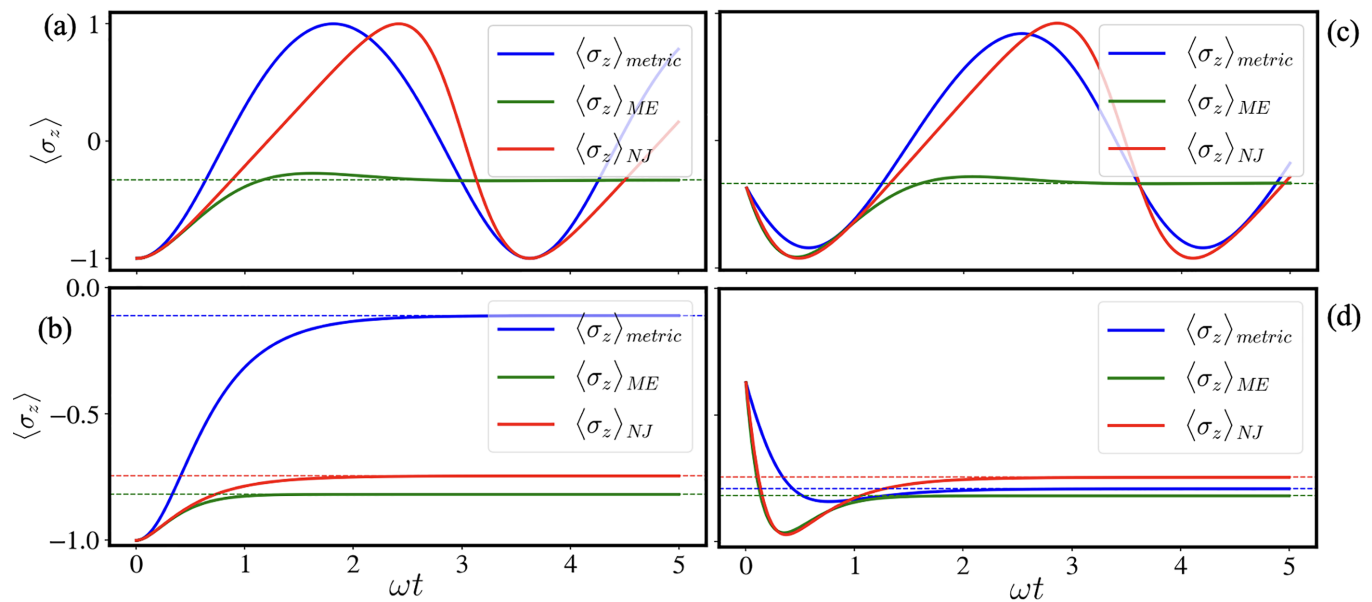

FIG. 1. The $\sigma_z$ expectation value for the model discussed in Sec. IV B, with (a),(c): $\mathcal{PT}$-symmetric regime $\omega = 1.0$, $\gamma = 0.5$ and (b),(d): $\mathcal{PT}$-broken regime $\omega = 1.0$, $\gamma = 1.5$. The solid lines show the three different approaches discussed in text: metric (blue; labeled $\langle\sigma_z\rangle_{\text{metric}}$), full master equation (green; labeled $\langle\sigma_z\rangle_{\text{ME}}$), and master equation in the no-jump limit (red; labeled $\langle\sigma_z\rangle_{\text{NJ}}$). The dashed lines with the same colors show the corresponding asymptotic values and/or steady-state solution shown in Eqs. (23)–(25). The initial conditions are (a),(b): the ground state of the $\sigma_z$ operator, and (c),(d): a random pure state $|\psi\rangle = (0.56 + 0.014i, -0.466 - 0.685i)^T$. The expectation values calculated with the no-jump approximation and the metric framework show parameter-dependent behavior. They are oscillatory and do not tend to a constant value for the parameter regime $\omega^2 > \gamma^2$ (a),(c). For $\omega^2 < \gamma^2$ (b),(d) where the energy eigenvalues of $H_{\text{eff}}$ are complex, the solutions approach different asymptotic values. The asymptotic value of the metric solution depends on the initial condition, which leads to a discrepancy in its relative value compared to the steady-state solution of the full master equation, as can be seen from (b) and (d). The qualitative behaviors of the no-jump solution and the full master equation solution are similar when $\omega^2 < \gamma^2$.

Unsurprisingly, the dependence of the metric solutions on the initial conditions leads to varying degrees of discrepancy with the full master equation solutions. For example, Fig. 1(b) shows a large discrepancy between $\langle\sigma_z\rangle_{\text{metric}}$ and $\langle\sigma_z\rangle_{\text{me}}$ in the long-time limit. However, this discrepancy is smaller in Fig. 1(d), where the asymptotic value of $\langle\sigma_z\rangle_{\text{metric}}$ is even closer to the steady-state solution of the master equation than the no-jump solution.

To summarize, the full master equation, the no-jump approximation and the metric approaches all display different behaviors in general, despite being model and parameter-specific. For the simpler model in Sec. IV A, the metric solution is oscillatory in general and does not tend to an asymptotic value in the long-time limit. On the other hand, the qualitative behavior of the no-jump solution resembles the master equation solution. In the long-time limit, the asymptotic value of the no-jump solution coincides with the steady-state solution of the master equation solution. The role of the exceptional points in the dynamics is evident in the more complicated model discussed in Sec. IV B. In the parameter regime where $\mathcal{PT}$ symmetry is preserved, there is a common qualitative feature between the metric and no-jump expectation values, which is oscillatory behavior. They only tend to an asymptotic value, resembling the steady-state solution, in the parameter regime where $\mathcal{PT}$ symmetry is spontaneously broken. These asymptotic values generally do not coincide with the steady-state solution of the master equation, though the no-jump approximation reproduces the qualitative behavior of the master equation albeit only in the $\mathcal{PT}$-broken regime. We also found that the asymptotic values of the metric expectation values are sensitive to initial conditions. This sensitivity can be attributed to Schrödinger evolution with an intrinsically non-Hermitian Hamiltonian, in contrast to the dissipative dynamics described by the full master equation. These observations motivate further study of whether there is a class of initial states where the metric formalism leads to observables which mimic steady-state values.

Our findings thus highlight the importance of considering the physical origin of the putative non-Hermiticity, since this dictates the choice of methodology needed to obtain the correct physical predictions. Namely, when non-Hermiticity is considered in the no-jump limit of a dissipative system, BQM and/or metric should not be used but instead the division-by-norm-method should be. This also demonstrates that a self-consistent non-Hermitian quantum system encapsulated by the metric formalism might harbor very different

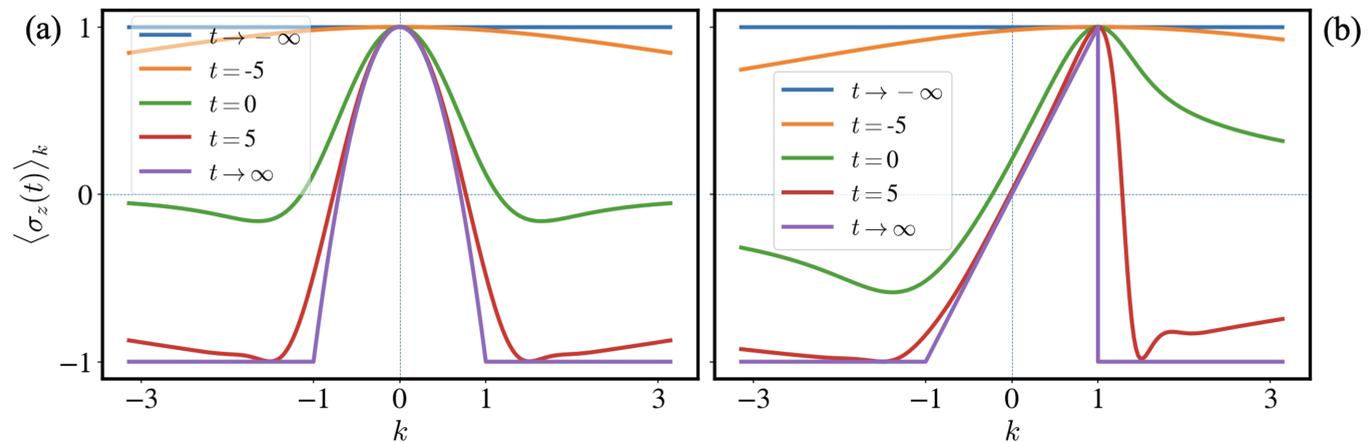

FIG. 2. The spin expectation value $\langle\sigma_z(t)\rangle_k$ at different times, calculated using (a) the metric framework, Eq. (10) and (b) the no-jump approximation (division by norm), Eq. (13). Each momentum mode $k$ is evolved using the non-Hermitian Hamiltonian given in Eq. (32) with $\gamma=1$ and $\Delta(t)=t$. The initial state at $t\to -\infty$ is $(1,0)^T$. (a) shows that $\langle\sigma_z\rangle_k$ is even in $k$ within the metric framework at all times. (b) shows that within the no-jump approximation, $\langle\sigma_z\rangle_k$ has no definite parity with respect to $k$ for finite times. In the asymptotic limit when $t\to\infty$, $\langle\sigma_z\rangle_k$ is an odd function of $k$ for the modes $|k|<\gamma$, which undergo spontaneous $\mathcal{PT}$-symmetry breaking during their evolution. However, for the modes with $|k|>\gamma$, which are the modes undergoing fully $\mathcal{PT}$-symmetric evolution, $\langle\sigma_z\rangle_k$ is an even function of $k$. This shows that when the metric dynamics is not taken into account, one fails to capture the underlying generalized transformation given by Eq. (28) for $\mathcal{PT}$-broken modes. For $|k|>\gamma$, the no-jump and the metric observables coincide in the asymptotic limit $t\to\infty$.

paradigms of physics compared to the standard no-jump limit of dissipative systems.

## V. GENERALIZED SIMILARITY TRANSFORMATIONS

In this section, we further explore the deep contrast between the consistent formalisms of non-Hermitian quantum mechanics, embodied by BQM and the metric, and the oft-used no-jump approximation valid for dissipative systems. Specifically, we focus on the properties of observables obtained using different methods. To this end, we generalize the notion of similarity transformation to self-consistent non-Hermitian quantum mechanics. A first point of significance in a non-Hermitian system is that the left and right eigenvectors of the Hamiltonian do not coincide, unlike in the Hermitian case. Therefore, the construction of transformations involving the eigenstates of the Hamiltonian should also take this aspect into account. The biorthogonal basis provides a natural path to this generalization. This notion is already naturally incorporated in the metric dynamics, which results in an important property of the observables within this framework.

This generalized similarity transformation also sheds light on the results of Ref. [49]. By incorporating the metric dynamics, it was found that spin expectation values possess fixed parity with respect to momentum at all times. This gives rise to profound physical impact in terms of defect production [49]. The origin of the evenness of $\langle\sigma_z\rangle_k$ with respect to momentum $k$, which was only observed in the metric framework, can be traced back to the presence of generalized similarity transformations of the Hamiltonian which maps the modes $k$ to $-k$. It is the construction of this transformation using both the left and right eigenvectors that gives rise to the evenness of $\langle\sigma_z\rangle_k$. Since the metric dynamics naturally incorporates this aspect in its construction, it mitigates the need to explicitly construct the transformation. In contrast, when physical quantities are simply divided by the time-dependent norm, corresponding to the no-jump approximation in dissipative systems, the spin expectation value does not have a definite parity with respect to $k\to -k$ at finite times. In the asymptotic limit $t\to\infty$, $\langle\sigma_z\rangle_k$ is even in $k$ for the modes which undergo fully $\mathcal{PT}$-symmetric time evolution. However, for the modes which undergo $\mathcal{PT}$-breaking during their evolution, $\langle\sigma_z\rangle_k$ is odd in $k$. These properties of the observable at various times are illustrated in Fig. 2.

We now turn to the explicit construction of the aforementioned similarity transformation using the left and right eigenstates. The spectral decomposition for a general non-Hermitian Hamiltonian, parametrized by a parameter $k$, acting on a finite-dimensional Hilbert space reads as follows:

$$H_k = \sum_n E_{k,n} |k,n\rangle_{RL}\langle k,n|. \quad (26)$$

For such a Hamiltonian, we can define the operators

$$U_{R,k} = \sum_n |-k,n\rangle_{RL}\langle k,n|,$$
$$U_{L,k} = \sum_n |-k,n\rangle_{LR}\langle k,n|, \quad (27)$$

which satisfy $U_{L,k} U_{R,k}^\dagger = U_{L,k}^\dagger U_{R,k} = \mathbb{1}$. This is the generalization of the notion of unitary transformations to a biorthogonal basis. We note that this is the first time such a generalization in the biorthogonal basis has been made, to the best of our knowledge.

These operators define the mappings between the eigenstates parametrized by $k$ and $-k$, which can be seen from $U_{R/L,k}|k,n\rangle_{R/L} = |-k,n\rangle_{R/L}$. We note that $U_{R,0} = U_{L,0} = \mathbb{1}$ for $k=0$, as the mapping is trivial in this case. If we impose the requirement that the spectrum is even with respect to $k$, such that $E_{k,n} = E_{-k,n}$ in Eq. (26), we have

$$U_{R,k} H_k U_{R,k}^{-1} = U_{R,k} H_k U_{L,k}^\dagger = H_{-k},$$
$$U_{L,k} H_k^\dagger U_{L,k}^{-1} = U_{L,k} H_k^\dagger U_{R,k}^\dagger = H_{-k}^\dagger. \quad (28)$$

This transformation of the Hamiltonian between the $k$ and $-k$ modes reflects the even parity of the spectrum of $H_k$.

We note that this left-right construction of mappings is not restricted to unitary transformations. In particular, this construction is also applicable to operations involving complex conjugation or transpose, which are known to be different for non-Hermitian Hamiltonians [59]. For example, we can demonstrate this by considering the case of the particle-hole symmetry (PHS) operators, $PHS$ and $PHS^\dagger$ considered in Ref. [59]. We take the operators $PHS := P$ and $PHS^\dagger := Q$ defined in Ref. [59] to be the right symmetry operators, such that

$$H = -P_R H^T P_R^{-1}, \quad H = -Q_R H^* Q_R^{-1}. \quad (29)$$

The corresponding left symmetry operators can then be defined as

$$H^\dagger = -P_L (H^\dagger)^T P_L^{-1}, \quad H^\dagger = -Q_L (H^\dagger)^* Q_L^{-1}. \quad (30)$$

We now compare Eq. (30) with the result obtained from the Hermitian conjugation of Eq. (29). This gives $P_L^{-1} = P_R^\dagger$ and $Q_R^{-1} = Q_L^\dagger$. The task of representing these operators in a structure similar to Eq. (27) merits further exploration.

A similar notion of similarity transformation also arises naturally within the metric framework. To this end, we can define a unitary operator similar to Eq. (27) using the spectrum of the Hermitian mapping $h_k$, given by Eq. (8). Writing $h_k = \sum_n e_{k,n} |k, n\rangle\langle k, n|$, the mapping is defined as

$$V_k = \sum_n |-k, n\rangle\langle k, n|. \quad (31)$$

It then follows that the even parity of the spectrum $e_k = e_{-k}$ translates directly to the transformation $V_k h_k V_k^\dagger = h_{-k}$. As stated in Sec. II B, $h_k$ and $H_k$ are equivalent representations of the same non-Hermitian system. Accordingly, the observables calculated using the evolution with $h_k$ and $H_k$ are the same within the metric framework, as shown in Eq. (10). Therefore, the observable calculated within the metric framework incorporates the mapping imposed by $V_k$, and thus $U_{R,k}$ and $U_{L,k}$. This shows that the notion of the generalized transformation between $H_k$ and $H_{-k}$ is already embedded in the metric framework, and we can access it directly in the observables without the need to consider the mapping given by Eq. (28).

To further explore this notion, we consider the concrete model of two level systems studied in Ref. [49],

$$H_k = k\sigma_x + i\gamma(t)\sigma_y + \Delta(t)\sigma_z, \quad (32)$$

where $k$ denotes momentum, as was considered in Ref. [49]. This Hamiltonian does not necessarily have to satisfy Eq. (14). In general, the coefficients $\gamma$ and $\Delta$ in Eq. (32) can be time dependent, in which case the operators defined in Eq. (27) should be constructed using the instantaneous eigenstates of the Hamiltonian. For a time-dependent Hamiltonian, the transformations are thus time-dependent in general.

As stated in the beginning of this section, Ref. [49] showed that the metric dynamics led to qualitatively different properties of spin expectation values. This was shown to have important physical consequences compared to the straightforward division by norm of the no-jump formalism. Here, we show that this property is a direct consequence of the generalized transformations discussed above. To this end, we construct the transformation operators for the Hamiltonian considered in Ref. [49], which corresponds to setting $\Delta(t) = Ft$ and $\gamma(t) = \gamma$ in Eq. (32), where $F$ and $\gamma$ are constant. The corresponding eigenvalues of the Hamiltonian are $E_k = \pm\sqrt{k^2 - \gamma^2 + \Delta^2}$, such that they satisfy the requirement for Eq. (27). The spectrum is real in the parameter regime $k^2 + \Delta^2 \geq \gamma^2$, where $\mathcal{PT}$ symmetry is not spontaneously broken, and forms complex conjugate pairs otherwise.

Using the instantaneous eigenstates of the Hamiltonian, we construct the operators $U_{L,k}$ and $U_{R,k}$ as in Eq. (27). Their exact forms are given by $U_{R,k} = U_{R,k}^\dagger = \frac{1}{\gamma+k}(\gamma \mathbb{1} - k\sigma_z)$ and $U_{L,k} = U_{L,k}^\dagger = \frac{1}{\gamma-k}(\gamma \mathbb{1} + k\sigma_z)$, such that Eq. (28) is satisfied. For this model, Eq. (31) greatly simplifies, giving $V_k = \sigma_z$. This implies the mapping $\sigma_z h_k(t) \sigma_z = h_{-k}(t)$.

In the metric framework, the expectation values are calculated using Eq. (10). Note that the normalized time-evolved state $|\Psi(t)\rangle_k$, as defined by Eq. (9), can be represented in the basis of the instantaneous eigenstates of $h_k(t)$ as $|\Psi(t)\rangle_k = \sum_{n \in \{\pm\}} c_{n,k}(t) |n(t)\rangle_k$. The aforementioned mapping $\sigma_z h_k(t) \sigma_z = h_{-k}(t)$ guarantees that the instantaneous eigenstates transform as $\sigma_z |n(t)\rangle_k \propto |n(t)\rangle_{-k}$. Additionally, for the choice of initial condition in Ref. [49] ($|\Psi(0)\rangle_k = (1, 0)^T$ for all $k$), the time-dependent coefficients of the time-evolved state are even in $k$, i.e., $c_{n,k}(t) = c_{n,-k}(t)$ for all $n$ at all times. Therefore, with this initial condition, the time-evolved state also transforms as $\sigma_z |\Psi(t)\rangle_k = |\Psi(t)\rangle_{-k}$.

Introducing the pure-state density matrix of the time-evolved state $\hat{\Psi}_k(t) \equiv |\Psi(t)\rangle_{k\,k}\langle\Psi(t)|$, it is easy to verify how the spin expectation values transform with respect to $k \to -k$, which is shown in the Supplemental Material of Ref. [49]. The density matrix is given by

$$\hat{\Psi}_k(t) = \frac{1}{2}\left(\mathbb{1} + \sum_{i=x,y,z} \langle\sigma_i(t)\rangle_k \sigma_i\right), \quad (33)$$

where $\langle\sigma_i(t)\rangle_k = \text{Tr}(\hat{\Psi}_k(t)\sigma_i)$. Since the density matrix transforms as $\sigma_z \hat{\Psi}_k \sigma_z = \hat{\Psi}_{-k}$ and $\sigma_z \hat{\Psi}_k \sigma_z = \frac{1}{2}(\mathbb{1} + \langle\sigma_z(t)\rangle_k \sigma_z - \sum_{i=x,y} \langle\sigma_i(t)\rangle_k \sigma_i)$, we see that $\langle\sigma_i(t)\rangle_k$ is even in $k$ for $i = z$ and odd in $k$ for $i = x, y$.

This transformation property of the observables is also visible in the BQM formalism. We expand Eq. (10) for a general observable $\hat{F}$ in the instantaneous biorthogonal basis

$$\langle\hat{F}\rangle_k = \text{Tr}(\hat{\Psi}_k(t)\hat{F})$$
$$= \sum_{ij \in \{\pm\}} {}_L\langle j|\hat{\Psi}_k(t)|i\rangle_R {}_L\langle i|\hat{F}|j\rangle_R, \quad (34)$$

where the time dependence in the basis has been omitted for notational ease. Since Eq. (27) implies that $|\pm\rangle_{R/L,-k} = U_{R/L,k}|\pm\rangle_{R/L,k}$, we have

$$\langle\hat{F}\rangle_{-k} = \sum_{m,n \in \{\pm\}} {}_{L,k}\langle m|U_{L,k}\sigma_z\hat{\Psi}_k\sigma_z U_{R,k}|n\rangle_{R,k}$$
$$\times {}_{L,k}\langle n|U_{L,k}\hat{F}U_{R,k}|m\rangle_{R,k}. \quad (35)$$

The first factor is independent of the observable $\hat{F}$, and can be recast as $U_{L,k}\sigma_z\hat{\Psi}_k\sigma_z U_{R,k} = \phi_{x,k}\sigma_x + \phi_{y,k}\sigma_y + \hat{\Psi}_k$. We can also rewrite the second term as $_{L,k}\langle n|U_{L,k}\hat{F}U_{R,k}|m\rangle_{R,k} \equiv c^F_{nm}\,_{L,k}\langle n|\hat{F}|m\rangle_{R,k}$, where the exact forms of the coefficients $\phi_{x/y,k}$ and $c^F_{nm}$ for $\hat{F} \in \{\sigma_x, \sigma_y, \sigma_z\}$ are summarized in Appendix B. Combining these results, we obtain

$$\langle F\rangle_{-k} = \sum_{m,n\in\{\pm\}} {}_{L,k}\langle m|\phi_{x,k}\sigma_x + \phi_{y,k}\sigma_y + \hat{\Psi}_k|n\rangle_{R,k}$$
$$\times\,_{L,k}\langle n|F|m\rangle_{R,k}c^F_{nm}. \quad (36)$$

Evaluating Eq. (36) for $\hat{F} = \sigma_z$ is algebraically simple, since the coefficient $c^{\sigma_z}_{nm} = 1$ for all pairs. We thus have

$$\langle\sigma_z\rangle_{-k} = \text{Tr}([\phi_{x,k}\sigma_x + \phi_{y,k}\sigma_y + \hat{\Psi}_k]\sigma_z) = \langle\sigma_z\rangle_k. \quad (37)$$

For the cases $F = \sigma_x, \sigma_y$, since the off-diagonal coefficients are nontrivial, a more complex calculation yields $\langle\sigma_i\rangle_{-k} = -\langle\sigma_i\rangle_k$ for $i = x, y$. In summary, we have introduced the generalization of similarity transformations using the biorthogonal basis, and showed how the mappings between the eigenstates of a non-Hermitian Hamiltonian with different momentum directly manifests in the metric formalism with important consequences for observables.

## VI. CONCLUSION

We have demonstrated the similarities and differences among the various methods available in the literature to study non-Hermitian systems. By employing illustrative models with exact solutions, we have shown that these methods yield not only quantitatively different results but also distinct physical interpretations. The physical origin of non-Hermiticity plays a crucial role in these differences.

Our findings indicate that when non-Hermiticity arises from dissipation in the limit where quantum jumps can be neglected, it is more appropriate to simply normalize physical quantities by dividing by the norm of the time-evolved state, especially in the $\mathcal{PT}$-broken regime. This method closely reproduces the qualitative behavior of the full master equation in this regime. Conversely, when non-Hermiticity is directly engineered in the system, such as by Naimark dilation, it is essential to consider the metric dynamics for a consistent quantum treatment. This approach can reveal richer physics, including the manifestation of a generalized transformation operator. Notably, our findings align with those of Refs. [60,61], which highlights starkly different behaviors obtained by BQM and division by the norm in the context of dynamical phase transitions in non-Hermitian systems.

Our work paves the way for deeper explorations of truly non-Hermitian Hamiltonians, where the physics of the metric and the concept of generalized transformations could potentially be harnessed for exotic behaviors. An exciting future direction is the extension of this framework to include other discrete and continuous symmetries. Although we would need other parameters to index different types of symmetries, it would still be necessary to include the left and right eigenstates in the construction of the transformations. This paves the way for the exploration of new paradigms for quantum criticality and spontaneous symmetry breaking.

## ACKNOWLEDGMENTS

K.S. is grateful for financial support from the Swiss National Science Foundation (SNSF) through Division II (No. 184739). This research is funded by the Deutsche Forschungsgemeinschaft (DFG, German Research Foundation) under Germany's Excellence Strategy EXC2181/1-390900948 (the Heidelberg STRUCTURES Excellence Cluster), the Swiss State Secretariat for Education, Research and Innovation (SERI), and by the Swedish Research Council (2018-00313 and 2024-05213) and Knut and Alice Wallenberg Foundation (KAW) via the project Dynamic Quantum Matter (2019.0068). The authors would like to thank G. M. Graf for numerous fruitful discussions.

## APPENDIX A: EXACT SOLUTION FOR THE METRIC APPROACH

In Sec. IV B, we discuss the case in which the jump operator $\Gamma^\dagger\Gamma$ does not commute with the system Hamiltonian. This leads to a rich dynamics which show qualitative differences using the different methods.

For the non-Hermitian Hamiltonian $H_{\text{eff}} = \omega\sigma_x - i\gamma\sigma_z$, the metric is given by

$$\rho(t) = \frac{\omega^2 - \gamma^2\cos(2\sqrt{\omega^2-\gamma^2}t)}{\omega^2 - \gamma^2}\mathbb{1}$$
$$- \frac{2\gamma\omega\sin^2(\sqrt{\omega^2-\gamma^2}t)}{\omega^2 - \gamma^2}\sigma_y + \frac{\gamma\sin(2\sqrt{\omega^2-\gamma^2}t)}{\sqrt{\omega^2-\gamma^2}}\sigma_z. \quad (A1)$$

This gives rise to the Hermitian mapping

$$h(t) = \left(2\omega + 2\gamma\frac{\rho_y(t)}{1 + \rho_0(t)}\right)\sigma_x \quad (A2)$$

given by Eq. (8).

## APPENDIX B: SPIN EXPECTATION VALUES

In Sec. V, we use the result $U_{L,k}\sigma_z\hat{\Psi}_k\sigma_z U_{R,k} = \phi_{x,k}\sigma_x + \phi_{y,k}\sigma_y + \hat{\Psi}_k$ to arrive at Eq. (36). Here, we provide the exact form of the coefficients

$$\phi_{x,k} = \frac{\gamma}{k^2 - \gamma^2}(\gamma\langle\sigma_x\rangle_k - ik\langle\sigma_y\rangle_k),$$
$$\phi_{y,k} = \frac{\gamma}{k^2 - \gamma^2}(\gamma\langle\sigma_y\rangle_k + ik\langle\sigma_x\rangle_k). \quad (B1)$$

TABLE I. The coefficients $c^F_{nm}$, defined as $_{L,k}\langle n|U_{L,k}FU_{R,k}|m\rangle_{R,k} \equiv c^F_{nm}\,_{L,k}\langle n|F|m\rangle_{R,k}$, where $F \in \{\sigma_x, \sigma_y, \sigma_z\}$ and $n, m = \pm$.

| $F$ | $\{+, +\}$ | $\{-, -\}$ | $\{+, -\}$ | $\{-, +\}$ |
|---|---|---|---|---|
| $\sigma_z$ | 1 | 1 | 1 | 1 |
| $\sigma_x$ | $-1$ | $-1$ | $\frac{\gamma E_k + k\Delta}{\gamma E_k - k\Delta}$ | $\frac{\gamma E_k - k\Delta}{\gamma E_k + k\Delta}$ |
| $\sigma_y$ | 1 | 1 | $\frac{\gamma\Delta + kE_k}{\gamma\Delta - kE_k}$ | $\frac{\gamma\Delta - kE_k}{\gamma\Delta + kE_k}$ |

We also rewrite the second term in Eq. (35) as $_{L,k}\langle n|U_{L,k}F U_{R,k}|m\rangle_{R,k} \equiv c^F_{nm}\,_{L,k}\langle n|F|m\rangle_{R,k}$. The coefficients $c^F_{nm}$ for $F \in \{\sigma_x, \sigma_y, \sigma_z\}$ and the pairs $\{n, m\}$ are summarized in Table I.

[1] Y. Ashida, Z. Gong, and M. Ueda, Non-Hermitian physics, Adv. Phys. **69**, 249 (2020).

[2] E. J. Bergholtz, J. C. Budich, and F. K. Kunst, Exceptional topology of non-Hermitian systems, Rev. Mod. Phys. **93**, 015005 (2021).

[3] C. M. Bender, Making sense of non-Hermitian Hamiltonians, Rep. Prog. Phys. **70**, 947 (2007).

[4] S. Sayyad and F. K. Kunst, Realizing exceptional points of any order in the presence of symmetry, Phys. Rev. Res. **4**, 023130 (2022).

[5] Z. Gong, Y. Ashida, K. Kawabata, K. Takasan, S. Higashikawa, and M. Ueda, Topological phases of non-Hermitian systems, Phys. Rev. X **8**, 031079 (2018).

[6] F. K. Kunst, E. Edvardsson, J. C. Budich, and E. J. Bergholtz, Biorthogonal bulk-boundary correspondence in non-Hermitian systems, Phys. Rev. Lett. **121**, 026808 (2018).

[7] F. Roccati, M. Bello, Z. Gong, M. Ueda, F. Ciccarello, A. Chenu, and A. Carollo, Hermitian and non-Hermitian topology from photon-mediated interactions, Nat. Commun. **15**, 2400 (2024).

[8] N. Matsumoto, K. Kawabata, Y. Ashida, S. Furukawa, and M. Ueda, Continuous phase transition without gap closing in non-Hermitian quantum many-body systems, Phys. Rev. Lett. **125**, 260601 (2020).

[9] N. Shibata and H. Katsura, Dissipative spin chain as a non-Hermitian Kitaev ladder, Phys. Rev. B **99**, 174303 (2019).

[10] T. Müller, S. Diehl, and M. Buchhold, Measurement-induced dark state phase transitions in long-ranged fermion systems, Phys. Rev. Lett. **128**, 010605 (2022).

[11] M. Buchhold, Y. Minoguchi, A. Altland, and S. Diehl, Effective theory for the measurement-induced phase transition of Dirac fermions, Phys. Rev. X **11**, 041004 (2021).

[12] Y. Wu, W. Liu, J. Geng, X. Song, X. Ye, C.-K. Duan, X. Rong, and J. Du, Observation of parity-time symmetry breaking in a single-spin system, Science **364**, 878 (2019).

[13] Y.-C. Chen, M. Gong, P. Xue, H.-D. Yuan, and C.-J. Zhang, Quantum deleting and cloning in a pseudo-unitary system, Front. Phys. **16**, 53601 (2021).

[14] C. Zheng, L. Hao, and G. L. Long, Observation of a fast evolution in a parity-time-symmetric system, Philos. Trans. R. Soc. A **371**, 20120053 (2013).

[15] J. M. Zeuner, M. C. Rechtsman, Y. Plotnik, Y. Lumer, S. Nolte, M. S. Rudner, M. Segev, and A. Szameit, Observation of a topological transition in the bulk of a non-Hermitian system, Phys. Rev. Lett. **115**, 040402 (2015).

[16] X. Zhang, Y. Tian, J.-H. Jiang, M.-H. Lu, and Y.-F. Chen, Observation of higher-order non-Hermitian skin effect, Nat. Commun. **12**, 5377 (2021).

[17] Q. Liang, D. Xie, Z. Dong, Haowei Li, Hang Li, H. Li, B. Gadway, W. Yi, and B. Yan, Dynamic signatures of non-Hermitian skin effect and topology in ultracold atoms, Phys. Rev. Lett. **129**, 070401 (2022).

[18] C. M. Bender and S. Boettcher, Real spectra in non-Hermitian Hamiltonians having $\mathcal{PT}$ symmetry, Phys. Rev. Lett. **80**, 5243 (1998).

[19] C. M. Bender, and D. W. Hook, $\mathcal{PT}$-symmetric quantum mechanics, Rev. Mod. Phys. **96**, 045002 (2024).

[20] C. M. Bender, S. Boettcher, and P. N. Meisinger, $\mathcal{PT}$-symmetric quantum mechanics, J. Math. Phys. **40**, 2201 (1999).

[21] J. Gong and Q.-h. Wang, Time-dependent $\mathcal{PT}$-symmetric quantum mechanics, J. Phys. A: Math. Theor. **46**, 485302 (2013).

[22] W. D. Heiss, The physics of exceptional points, J. Phys. A: Math. Theor. **45**, 444016 (2012).

[23] C. H. Lee, Exceptional bound states and negative entanglement entropy, Phys. Rev. Lett. **128**, 010402 (2022).

[24] L. Ding, K. Shi, Q. Zhang, D. Shen, X. Zhang, and W. Zhang, Experimental determination of $\mathcal{PT}$-symmetric exceptional points in a single trapped ion, Phys. Rev. Lett. **126**, 083604 (2021).

[25] W. Chen, Ş. Kaya Özdemir, G. Zhao, J. Wiersig, and L. Yang, Exceptional points enhance sensing in an optical microcavity, Nature (London) **548**, 192 (2017).

[26] J. C. Budich and E. J. Bergholtz, Non-Hermitian topological sensors, Phys. Rev. Lett. **125**, 180403 (2020).

[27] B. Dóra and C. P. Moca, Quantum quench in $\mathcal{PT}$-symmetric Luttinger liquid, Phys. Rev. Lett. **124**, 136802 (2020).

[28] D. C. Brody, Biorthogonal quantum mechanics, J. Phys. A: Math. Theor. **47**, 035305 (2014).

[29] T. Curtright and L. Mezincescu, Biorthogonal quantum systems, J. Math. Phys. **48**, 092106 (2007).

[30] J. V. Hounguevou, F. A. Dossa, and G. Y. H. Avossevou, Biorthogonal quantum mechanics for non-Hermitian multimode and multiphoton Jaynes–Cummings models, Theor. Math. Phys. **193**, 1464 (2017).

[31] A. Felski, A. Beygi, C. Karapoulitidis, and S. P. Klevansky, Three perspectives on entropy dynamics in a non-Hermitian two-state system, Phys. Scr. **99**, 125234 (2024).

[32] H. B. Geyer, W. D. Heiss, and F. G. Scholtz, The physical interpretation of non-Hermitian Hamiltonians and other observables, Can. J. Phys. **86**, 1195 (2008).

[33] A. Mostafazadeh, Time-dependent pseudo-Hermitian Hamiltonians and a hidden geometric aspect of quantum mechanics, Entropy **22**, 471 (2020).

[34] A. Mostafazadeh, Energy observable for a quantum system with a dynamical Hilbert space and a global geometric extension of quantum theory, Phys. Rev. D **98**, 046022 (2018).

[35] H. Badhani and S. Ghosh, Decomposition of a system in pseudo-Hermitian quantum mechanics, J. Phys. A Math. Theor. (2025), doi:10.1088/1751-8121/adc216.

[36] X. Turkeshi and M. Schiró, Entanglement and correlation spreading in non-Hermitian spin chains, Phys. Rev. B **107**, L020403 (2023).

[37] B. Dóra, M. Heyl, and R. Moessner, The Kibble-Zurek mechanism at exceptional points, Nat. Commun. **10**, 2254 (2019).

[38] A. Bácsi and B. Dóra, Dynamics of entanglement after exceptional quantum quench, Phys. Rev. B **103**, 085137 (2021).

[39] B. Dóra, D. Sticlet, and C. P. Moca, Correlations at PT-symmetric quantum critical point, Phys. Rev. Lett. **128**, 146804 (2022).

[40] B. Longstaff and E.-M. Graefe, Nonadiabatic transitions through exceptional points in the band structure of a *PT*-symmetric lattice, Phys. Rev. A **100**, 052119 (2019).

[41] I. Percival, *Quantum State Diffusion* (Cambridge University Press, Cambridge, 1999).

[42] H. M. Wiseman and G. J. Milburn, Quantum measurement theory, in *Quantum Measurement and Control* (Cambridge University Press, Cambridge, 2009), pp. 1–50.

[43] T. A. Brun, A simple model of quantum trajectories, Am. J. Phys. **70**, 719 (2002).

[44] R. D. Soares and M. Schirò, Non-unitary quantum many-body dynamics using the Faber polynomial method, SciPost Phys. **17**, 128 (2024).

[45] M. Stefanini and J. Marino, Orthogonality catastrophe beyond bosonization from post-selection, Phys. Rev. Res. **6**, L042022 (2024).

[46] H.-P. Breuer and F. Petruccione, *The Theory of Open Quantum Systems* (Oxford University Press, Oxford, 2002).

[47] C. M. Bender, D. C. Brody, H. F. Jones, and B. K. Meister, Faster than Hermitian quantum mechanics, Phys. Rev. Lett. **98**, 040403 (2007).

[48] A. Mostafazadeh, Quantum brachistochrone problem and the geometry of the state space in pseudo-Hermitian quantum mechanics, Phys. Rev. Lett. **99**, 130502 (2007).

[49] K. Sim, N. Defenu, P. Molignini, and R. Chitra, Quantum metric unveils defect freezing in non-Hermitian systems, Phys. Rev. Lett. **131**, 156501 (2023).

[50] F. Bagarello, Transition probabilities for non self-adjoint Hamiltonians in infinite dimensional Hilbert spaces, Ann. Phys. **362**, 424 (2015).

[51] A. Mostafazadeh, Pseudo-hermiticity and generalized PT- and CPT-symmetries, J. Math. Phys. **44**, 974 (2003).

[52] A. Mostafazadeh, Time-dependent Hilbert spaces, geometric phases, and general covariance in quantum mechanics, Phys. Lett. A **320**, 375 (2004).

[53] G. Lindblad, On the generators of quantum dynamical semigroups, Commun. Math. Phys. **48**, 119 (1976).

[54] B. Gulácsi and B. Dóra, Defect production due to time-dependent coupling to environment in the Lindblad equation, Phys. Rev. B **103**, 205153 (2021).

[55] M. Coppola, D. Karevski, and G. T. Landi, Conditional no-jump dynamics of noninteracting quantum chains, Phys. Rev. B **110**, 094315 (2024).

[56] A. Mostafazadeh, Pseudo-Hermitian representation of quantum mechanics, IJGMMP **07**, 1191 (2010).

[57] A. Mostafazadeh, Pseudo-Hermiticity versus PT symmetry: The necessary condition for the reality of the spectrum of a non-Hermitian Hamiltonian, J. Math. Phys. **43**, 205 (2002).

[58] A. Das, Pseudo-Hermitian quantum mechanics, J. Phys.: Conf. Ser. **287**, 012002 (2011).

[59] K. Kawabata, K. Shiozaki, M. Ueda, and M. Sato, Symmetry and topology in non-Hermitian physics, Phys. Rev. X **9**, 041015 (2019).

[60] Y. Jing, J.-J. Dong, Y.-Y. Zhang, and Z.-X. Hu, Biorthogonal dynamical quantum phase transitions in non-Hermitian systems, Phys. Rev. Lett. **132**, 220402 (2024).

[61] H. Badhani, S. Banerjee, and C. M. Chandrashekar, Non-Hermitian quantum walks and non-Markovianity: The coin-position interaction, Phys. Scr. **99**, 105112 (2024).